\documentclass[twocolumn,showpacs]{revtex4}
\usepackage{graphicx}
\usepackage{dcolumn}
\usepackage{bm}

\begin{document}

\title{Stochastically sustained population oscillations in high-$\beta$ nanolasers}

\author{A. Lebreton}
\author{I. Abram}
\affiliation{Laboratoire de Photonique et Nanostructures LPN-CNRS UPR20, Route de Nozay, 91460 Marcoussis, France}

\author{N. Takemura}
\affiliation{Department of Physics, The University of Tokyo, Hongo, Tokyo 113-0033, Japan}

\author{M. Kuwata-Gonokami}
\affiliation{Department of Physics, The University of Tokyo, Hongo, Tokyo 113-0033, Japan}
\affiliation{Photon Science Center, The University of Tokyo, Hongo, Tokyo 113-0056, Japan}

\author{I. Robert-Philip}
\email{isabelle.robert@lpn.cnrs.fr}
\author{A. Beveratos}

\affiliation{Laboratoire de Photonique et Nanostructures LPN-CNRS UPR20, Route de Nozay, 91460 Marcoussis, France}


\begin{abstract}
Non-linear dynamical systems involving small populations of individuals may sustain oscillations in the population densities arising from the discrete changes in population numbers due to random events. 
By applying these ideas to nanolasers operating with small numbers of emitting dipoles and photons at threshold, we show that such lasers should display photon and dipole population cycles above threshold, which should be observable as a periodic modulation in the second-order correlation function of the nanolaser output. Such a modulation was recently reported in a single-mode vertical-cavity surface-emitting semiconductor laser. 
\end{abstract}
\maketitle
\section{Introduction}

Nanolasers, such as semiconductor microcavity lasers \cite{Strauf2006, Ulrich2007, Hostein2010} or plasmonic lasers \cite{Noginov2009, Hill2009, Nezhad2010}, display cavity volumes of the order of a fraction of a cubic wavelength and thus contains few emitters, while their lasing threshold is attained with very few photons in the cavity, of the order of $\beta^{-1/2}$ \cite{Rice} at any given time, where $\beta$ is the fraction of spontaneous emission funnelled in the ``useful'' mode. 
In a nanocavity, $\beta$ may be close to 1, so that the number of photons at threshold is very small, typically 1 to 10.
For such small numbers, fluctuations are large (of the order of the average values), occur only in integral multiples of a base value (corresponding to a change of a single unit), are asymmetric (as excursions into negative values are impossible) and cause the nanolaser output to deviate from that predicted by the traditional laser rate equations \cite{Choudhury, Elvira}. 
The reason for this deviation is that these real-number differential equations have been developed for conventional macroscopic lasers, in which the discrete nature of the number of emitters and photons may be ignored (termed ``the thermodynamic limit'' \cite{Rice}) and are therefore inadequate for describing the operation of nanolasers.

Theoretical methods based on full quantum electrodynamic treatments have been developed \cite{Gardiner} to account for phenomena observed in single-atom or few-atom microlasers and micromasers, which have been studied experimentally since the 1980s \cite{Haroche, Walther, Kimble}. 
However, in these systems the emitting atoms are relatively well isolated from their environment, have sharp optical spectra, and generally couple strongly to the cavity mode. 
Semiconductor (or plasmonic) nanolasers, on the other hand, operate at the opposite limit: The emitters (generally semiconductor quantum dots) are embedded in the cavity material and thus their radiating dipoles interact extensively with their environment, undergoing very fast dephasing processes. At the same time, under the strong pumping conditions required for lasing, the emitters are usually highly-excited, containing several electron-hole pairs (dipoles) and producing very broad emission spectra \cite{Winger}.
These conditions are contrary to the assumptions of the few-atom microlaser theories, as they preclude strong coupling between the emitters and the cavity mode, while the incoherent fluctuation processes cause decay of the atomic polarization  and rapidly wash out any phase-related phenomena.
In this limit, the full quantum equations reduce to the traditional laser equations that involve only photon and dipole numbers \cite{Rice}.
However, these are continuous-variable differential equations that describe the evolution of the mean photon and dipole populations, but ignore their discrete nature: 
In each individual realization of the lasing process only integer values of photon and dipole numbers may be obtained, a feature that may be ignored in large lasers but becomes important in nanolasers where these numbers are small.
As described below, the traditional laser equations can be adapted to the discrete nature of the photon and dipole populations through discrete simulation methods in which small increments in time are made, and within each increment a discrete but random number of constituents is chosen, obeying the statistics of the physical process being modelled. 

This procedure is quite similar to that used in modelling small predator-prey or epidemiological ecosystems \cite{McKane2005, Aparicio2001}, where it has been shown that ``demographic stochasticity'', that is population fluctuations due to random birth, death, predation or contagion events, produces cyclical population variations.
These cycles are due to the discrete nature of the individuals in the population and cannot be accounted for through traditional deterministic models for the populations, based on real-number continuous differential equations, such as the Lotka-Volterra equations, whose domain of validity corresponds to the cases in which discretization may be ignored, such as in infinite ecosystems, or when extensive ensemble-averaging is possible. 

The traditional method of dealing with fluctuations in lasers consists of inserting randomly-fluctuating Langevin forces in the laser rate equations and linearizing about the steady-state photon and dipole populations \cite{Druten2000}.
However, this method is not suitable when small populations are involved, since in that case the magnitude of the fluctuations is of the same order as the mean values (and thus linearization is not an adequate approximation) while, at the same time, the small-population statistics deviate significantly from the Gaussian statistics of the Langevin forces. 
Thus, while the traditional treatment may be adjusted so that the variance of the Langevin forces reproduces the observed variance of the fluctuations, it cannot account adequately for the higher-order moments of their distribution.
In a recent publication \cite{Elvira} we measured the higher-order intensity fluctuation statistics of a pulsed nanolaser (up to the fourth order) and, through the implementation of a discrete simulation method, we showed that the discrete nature of the photon number at threshold induces a jitter in the pulse timing, which accounts for the observed photon statistics to all orders.

In this paper, we examine a continuously pumped nanolaser and show that the ``granularity'' of the photon and dipole populations and the ``discretization noise" it introduces lead to large self-sustained oscillations in these populations, often reaching extinction periodically for the case of photons, even above threshold. 
While this behavior is similar to that of small predator-prey ecosystems, it can also be understood within the traditional laser theory as arising from the laser's relaxation oscillations driven by the discrete jumps associated with the spontaneous and stimulated emission events. 
The photon population cycles in nanolasers should be observable as a periodic modulation of $g^{(2)}(\tau)$, the intensity autocorrelation function of the nanolaser.

Such a periodic modulation has recently been reported by Takemura {\it et al.} \cite{Takemura} in experiments involving a single-mode vertical-cavity surface-emitting laser (VCSEL). While VCSELs are not nanolasers, the sample examined displayed $\beta \approx 10^{-4}$, corresponding to $\beta^{-1/2} \approx 100$ photons in the cavity at threshold, so that a periodic modulation of  $g^{(2)}(\tau)$, albeit of small amplitude, could be measured in the vicinity of the threshold.
The discretization noise model accounts well for this modulation, thus validating its use in solid-state nanolasers.

\section{Discretization noise}

We consider an optical cavity with one only ``useful'' (or reference) mode, while all other modes are considered as ``leaks''. 
The cavity contains an ensemble of $N$ excited emitters (dipoles) each emitting photons at a spontaneous rate $\gamma_\parallel$.
We assume that the emitters interact extensively with their material environment, so that they undergo very rapid dephasing processes (as is the case, for example, in solid-state room-temperature lasers) in a time short compared with the characteristic time of the coupling with the electromagnetic field.

Under these assumptions, the quantum mechanical Maxwell-Bloch equations (or, equivalently, the Jaynes-Cummings model), which describe the dipole-field interaction, may be simplified by adiabatically eliminating the emitter polarization, so that the emission kinetics in such a system is adequately described by the traditional laser rate equations \cite{Rice}:
\begin{eqnarray} 
\label{eq:Rate1}
\frac{dN(t)}{dt} = & P(t) -\gamma_\parallel N(t) & -\beta \gamma_\parallel N(t) s(t) \\ 
\label{eq:Rate2}
\frac{ds(t)}{dt} = & -\Gamma_c s(t) +\beta \gamma_\parallel N(t) &+ \beta \gamma_\parallel N(t) s(t)
\end{eqnarray}
\noindent where $P(t)$ is the incident pumping rate; $s(t)$ is the number of photons in the lasing (reference) mode; $1/\Gamma_c$ and $1/\gamma_\parallel$ are respectively the lifetime of the reference mode (also termed cavity lifetime) and the dipole lifetime; and $\beta$ is the fraction  of spontaneous emission funnelled into the reference mode.

These are continuous-variable differential equations that describe the evolution of the mean photon and dipole populations, but ignore the ``granularity'', that is the discrete nature of these populations.
In any particular realization of the lasing process, the photon and dipole populations can assume only integer values, with discrete transitions between the different population configurations. The physical processes behind these transitions are the excitation of a dipole through ``pumping'', the emission of a photon by a dipole inside or outside the reference mode, or the escape of a photon from the reference mode to the outside.
The discrete nature of the photon and dipole numbers can be taken into account rigorously (and under the same assumptions of rapid loss of phase memory) through the master equation approach, which considers the probability of each realization of the discrete photon and dipole populations, as well as the transitions between the different population configurations  \cite{Rice}. 
It can therefore account for the full statistics of the emission process.
Within the master equation approach the discrete transitions between different population configurations can be regarded as probabilistic partitions of the photon and dipole numbers produced by the processes of pumping, emission, and escape, governed by the statistical parameters of the master equation.

When the master equation is applied to the calculation of the mean photon and dipole numbers ($\langle s \rangle$ and $ \langle N \rangle $), it yields the traditional laser rate equations (\ref{eq:Rate1}) and (\ref{eq:Rate2}), under the assumption that the fluctuations in the dipole and photon numbers are uncorrelated ($\langle N s \rangle = \langle N \rangle \langle s \rangle $) \cite{Rice, Choudhury}, which is a good approximation when the number of photons and dipoles is large (so that their relative fluctuations are small) or, more precisely, in the ``thermodynamic limit'', that is when $1/\beta \rightarrow \infty$.

Calculation of the laser photon statistics by use of the master equation requires calculating at each time-step a matrix of $(N_{max} \times s_{max})$ interdependent probabilities (where the subscript ``$max$'' indicates a cut-off value chosen large enough so as to encompass all population values with a significant probability), a formidable task when the dimensionality of the matrix is larger than a few tens.
Alternatively, the master equation may be solved through stochastic simulation methods \cite{Choudhury, McKane2005}, whereby the system evolves from one discrete state to another through a jump event in each time-step, thus describing a trajectory in state-space.
The average of a large number of trajectories converges to the solution of the master equation.
This greatly simplifies the calculation, as each trajectory is calculated independently of the others, and convergence of the average is reasonably rapid.
Beyond the mathematical convenience it offers in the calculation of the master equation, the calculation of individual trajectories permits a direct visualization of the time-evolution of the system parameters and thus provides physical insight into the ongoing processes.

The laser rate equations provide a guideline in the calculation of the stochastic state-space trajectories, since these equations deal with the mean values that may be calculated through the master equation.
Thus, at each step, the mean of the stochastic evolution of a given initial state must follow the laser rate equations. 

To this end, the rate equations  (\ref{eq:Rate1}) and (\ref{eq:Rate2}) can be solved numerically by combining an iterative procedure over a time-step $dt$, small compared with $1/\Gamma_c$ and $1/\gamma_{\parallel}$, and a quantum jump approach to ensure that $N$ and $s$ are integers. The use of a small time-step ensures that this iterative procedure follows well the fine structure of the time-evolution of the system, while at the same time it preserves the correlations between the fluctuations in the photon and dipole populations.
Thus, re-writing the laser rate equations, the discrete evolution of $N$ and $s$ follows at each iteration step:
\begin{eqnarray}
\label{eq:Rate3}
N(t+dt) = & N(t) + P(t)dt - \gamma_{\parallel} (1+\beta s(t))  N(t)dt \\ 
\label{eq:Rate4}
s(t+dt) =& s(t) -\Gamma_c s(t)dt + (1+s(t))\beta \gamma_{\parallel} N(t)dt
\end{eqnarray}
\noindent 
This presentation of the laser rate equations permits us to interpret each increment or decrement of $N$ and $s$ as resulting from a random drawing from an appropriate distribution, such that its average is the corresponding expression in these equations.
Thus, denoting by $PD[\mu]$ a random number out of a Poisson distribution with mean value $\mu$, and by $BD[n,p]$ a random number drawn from a binomial distribution of $n$ trials, each of which yields success with probability $p$, we can obtain the number of additional dipoles excited by the pumping process during time-step $dt$ as $PD[P(t) dt]$, while $N_d(t)$, the number of dipoles having decayed between $t$ and $t + dt$ by spontaneous or stimulated emission, is given by $N_d(t)$=$BD[N(t),\gamma_{\parallel}(1+\beta s(t)) dt]$.  
Each photon emitted by these dipoles has a probability $(1+s(t))\beta/(1+\beta s(t))$ to be funneled in the cavity mode, and thus the number of emitted photons entering the cavity mode is $BD[N_d(t), (1+s(t))\beta/(1+\beta s(t))]$.
Finally, the number of photons that escape to the outside due to cavity losses, is drawn from the binomial distribution $BD[s(t),\Gamma_c dt]$.

This procedure accounts for the evolution of the photon and dipole populations as integer numbers. It thus takes account of the ``discretization noise'' without any need to introduce it in the continuous-variable laser rate equations externally.
Traditionally, noise (including quantum fluctuations) is re-introduced in the laser equations (\ref{eq:Rate1}) and (\ref{eq:Rate2}) by linearizing around a (large) mean value of the populations and adding a Langevin driving force with Gaussian statistics \cite{Druten2000}. 
However, this approach is valid only in the case of large populations \cite{McKane2005}, where the relative fluctuations are small and obey Gaussian statistics thanks to the central limit theorem.

\begin{figure*}[!ht]
   \begin{center}
     
   \includegraphics[width=13.75cm]{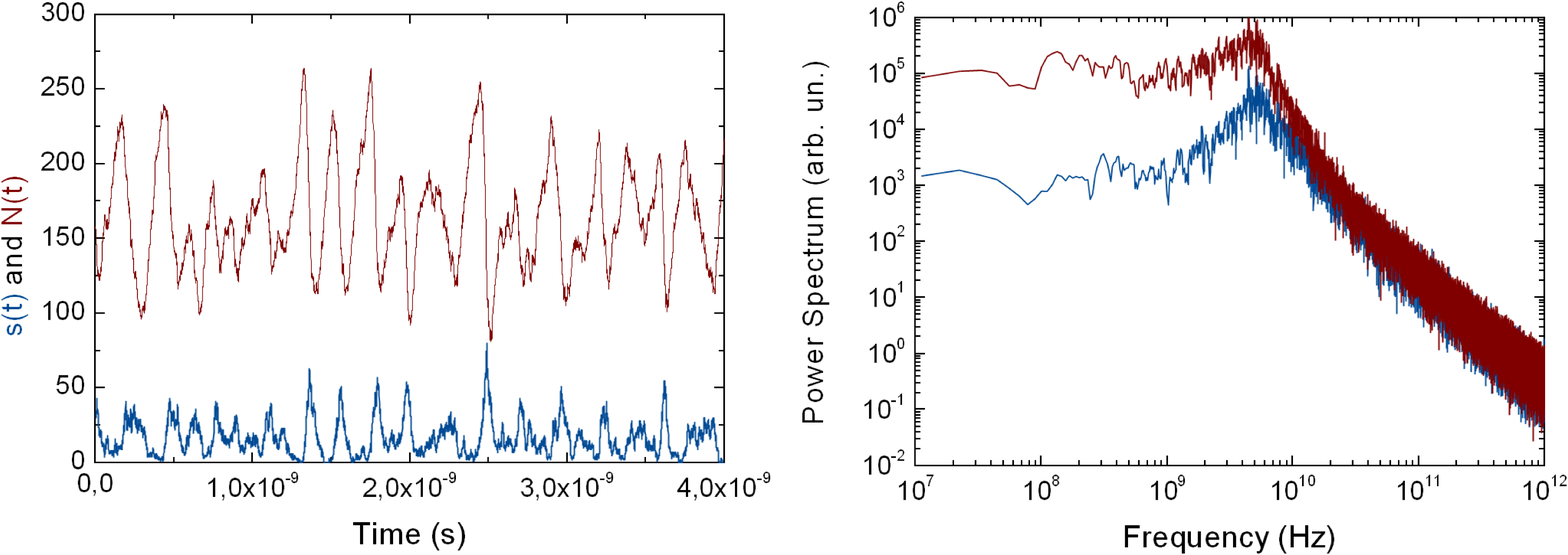} 
   
   \end{center}
   \caption{ \label{fig:cycles-freq} (Left) Evolution of the number of photons $s(t)$ (blue line) and the number of excited dipoles $N(t)$ (red line) as a function of time for a nanolaser with $\beta = 0.25$, $1/\gamma_\parallel = 400$ ps and $1/\Gamma_c = 10$ ps for a pumping power of $P = 4.5 P_{th}$, with $P_{th} =\Gamma_c/\beta$, the nominal threshold pumping power.
(Right) The corresponding power spectra.}
\end{figure*}

\section{Population oscillations in nanolasers}

To gain an insight into the effects of discretization noise, we have calculated state-space trajectories at different pumping powers, by solving Equations (\ref{eq:Rate3}) and (\ref{eq:Rate4}) numerically through the above procedure, using the values of $\beta=0.25$, $1/\gamma_\parallel = 400$ ps and $1/\Gamma_c = 10$ ps (corresponding to a quality factor of 13000 at $\lambda = 1.5 \mu$m), which are typical for quantum dot nanolasers \cite{Strauf2006, Ulrich2007, Hostein2010}.

\begin{figure*}[!ht]
   \begin{center}
    \begin{center}
     
   \includegraphics[width=13.75cm]{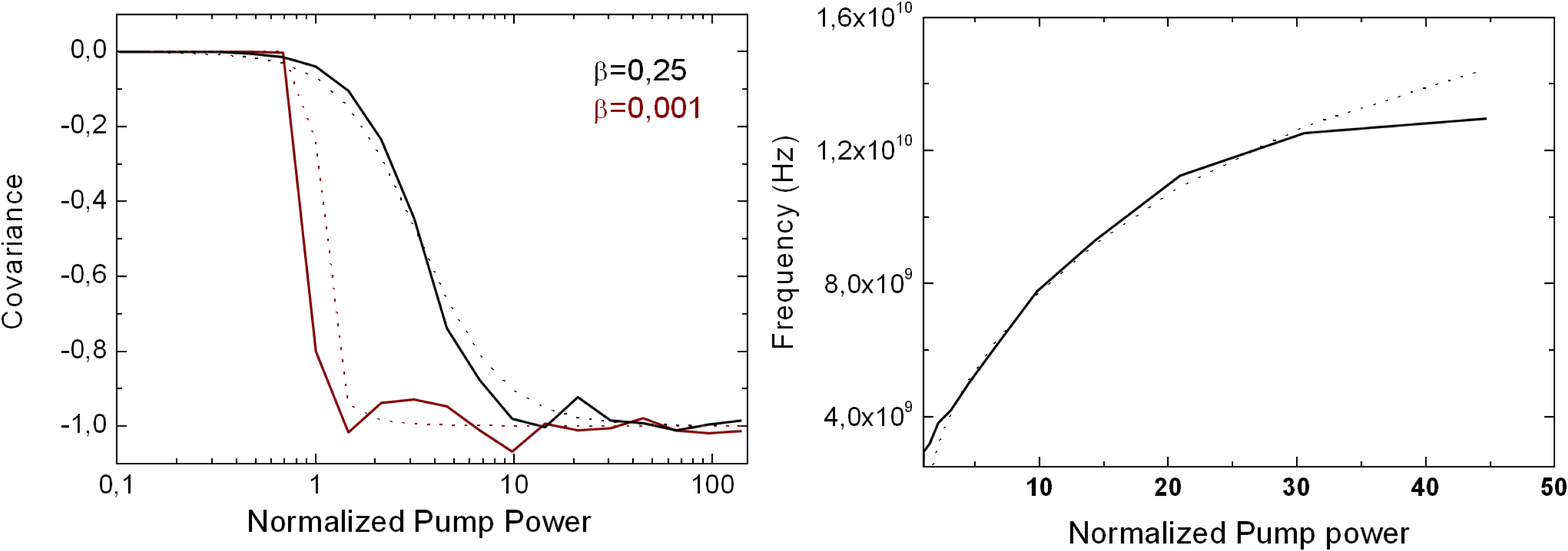} 
   
   \end{center}
   \end{center}
   \caption{ \label{fig:covar-relax-oscil} (Left) Covariance of the number of photons $s(t)$ and the number of excited dipoles $N(t)$ as a function of the pumping power, for $\beta=0.25$ (black line) and $\beta=0.001$ (red line). 
   Dotted curves are calculated through the relaxation oscillation theory. 
   Curves are rescaled by the pinning value of the dipoles $\Gamma_c/\beta \gamma_\parallel$.
   (Right) Comparison of the photon population cycle frequency for $\beta=0.25$ (continuous line) with the relaxation oscillation frequency (dotted line), as a function of pumping power.}
\end{figure*}

Figure 1 (left) displays the time-dependence of the number of intra-cavity photons and excited dipoles at a pumping power of $4.5 P_{th}$, where $P_{th}=\Gamma_c/\beta$ is the nominal threshold pumping power.
At this pumping power, the mean number of dipoles has reached its pinning value of $\Gamma_c/\beta \gamma_\parallel \approx 160$, yet the photon and dipole populations display well-defined oscillations of the same frequency, as can be seen in the corresponding power spectra (Fig.\ 1 right). These oscillations are synchronized (but in quadrature), in a manner analogous to the population cycles driven by demographic stochasticity observed in finite predator-prey ecosystems \cite{McKane2005}. 

The synchronization of the photon and dipole population cycles as a function of pumping power can be studied through the covariance of these two quantities, as can be seen in Fig.\ \ref{fig:covar-relax-oscil} (left) (black continuous line), where the covariance curve for a low-$\beta$ laser (red continuous line) is also given for comparison. 
Below threshold, the photon and dipole populations fluctuate randomly and independently of each other, as the presence of less than one photon (on average) in the cavity produces no change in the dipole emission dynamics and spontaneous emission outside the cavity mode destroys any correlation between these two populations. 
At higher powers, as stimulated emission grows, a larger fraction of the emission goes into the reference mode, so that any decrease in the dipole population is matched by a synchronous increase of the intra-cavity photon population, thus producing a strong negative correlation.
The difference between high- and low-$\beta$ lasers is essentially the speed with which the rescaled covariance reaches its limiting value of -1.
At a given pumping power (normalized with respect to threshold), a high-$\beta$ laser has a smaller number of intra-cavity photons, so that its stimulated emission rate is smaller implying that a smaller fraction of the emission goes into the reference mode. 
As a consequence, the rescaled covariance drops more gradually to its limiting value of -1.

The power dependance of the covariance (Fig.\ \ref{fig:covar-relax-oscil} left) and of the central frequency of the population oscillations (Fig.\ \ref{fig:covar-relax-oscil} right) are quite similar to those calculated through the traditional linearized ocillation relaxation theory \cite{Henry} (Fig.\ \ref{fig:covar-relax-oscil} dotted lines), indicating that the photon and dipole population cycles in a nanolaser can be understood also as arising from relaxation oscillations driven by the noise due to the granularity of photon emission. This mechanism is similar to that of the origin of the excess linewidth in semiconductor lasers \cite{Henry} and of the excess noise in class-B (i.e.\ $\Gamma_c > \gamma_\parallel$) lasers \cite{Druten2000}, however with an important difference. 
In traditional low-$\beta$ lasers, the (spontaneous and stimulated) emission events are distributed essentially uniformly in time, corresponding to a white noise spectrum that gets filtered by the transfer function of the system to produce a damped oscillation. 
On the other hand, in high-$\beta$ lasers, because of the small mean value of the intra-cavity photons, emission events are more frequent at the top of the photon population cycle, and thus constitute a periodic driving force that produces a resonant amplification sustaining the photon and dipole population cycles. 

\begin{figure}[!ht]
   \begin{center}
      \begin{tabular}{l}
   \includegraphics[width=7cm]{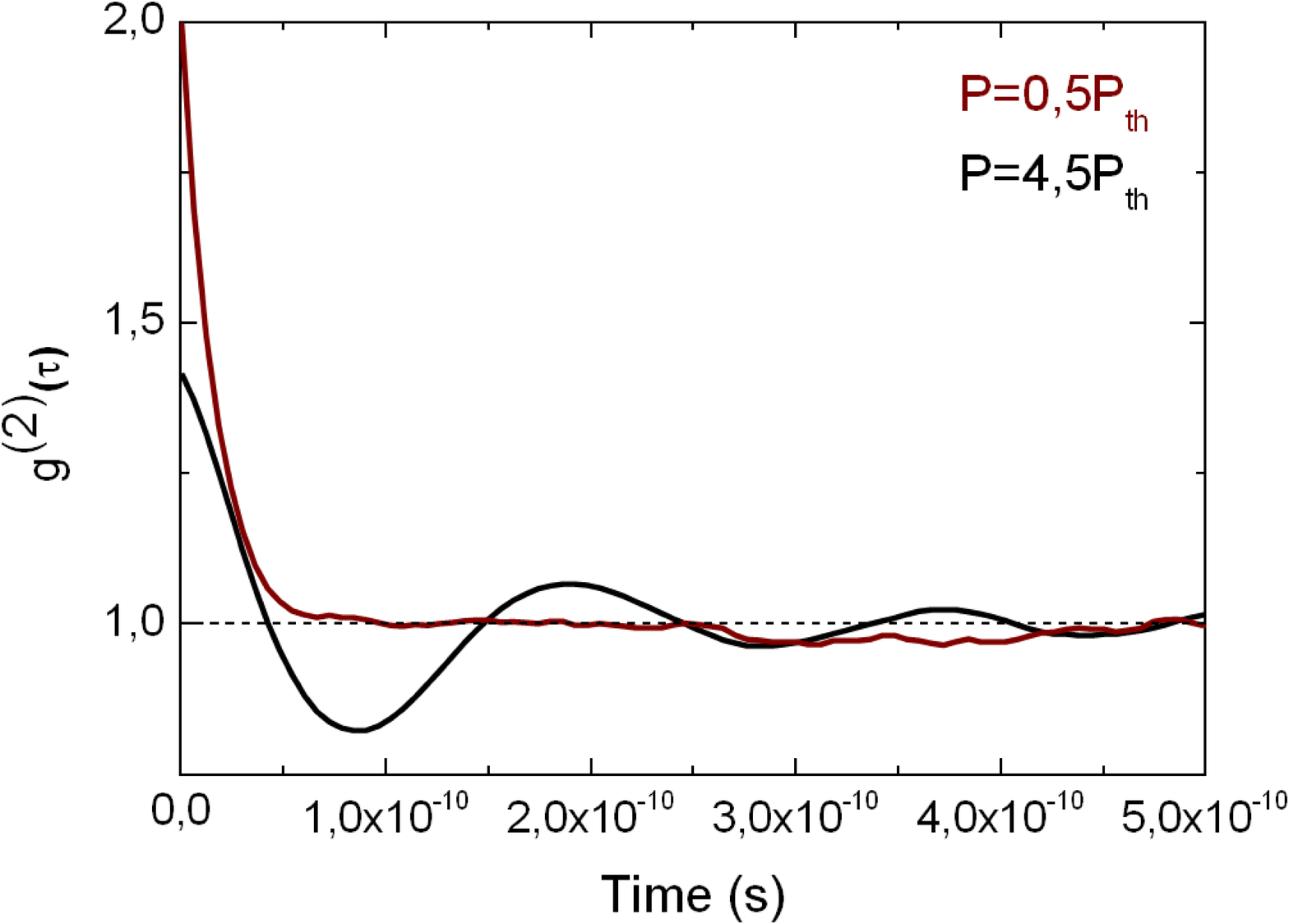} 
     \end{tabular}
   \end{center}
   \caption{ \label{fig:g2} Second-order photon correlation function $g^{(2)}(\tau)$ of a nanolaser ($\beta=0.25$) undergoing photon population cycles, for $P=0.5 P_{th}$ (red line) and $P=4.5 P_{th}$ (black line).
The dashed horizontal line corresponds to $g^{(2)}(\tau)=1$, the value in continuous-wave lasers. }
\end{figure}

The photon population cycles should be directly observable in the second-order photon correlation function $g^{(2)}(\tau)$, which is measurable experimentally through a Hanbury Brown and Twiss setup. 
As can be seen on Fig.\ \ref{fig:g2} (red line), below threshold $g^{(2)}(\tau)$ displays the usual behavior for chaotic light. 
That is, when the two detected photons are in temporal coincidence it has the value $g^{(2)}(0)=2$, while when the two photons are separated by a time interval $\tau$ the value of $g^{(2)}(\tau)$ drops to 1 exponentially $e^{-\Gamma_c \tau}$, where $\Gamma_c$ is the cavity energy decay rate. This is the familiar Siegert relation, 
\begin{equation}
g^{(2)}(\tau)=1+ \left | g^{(1)}(\tau) \right |^2 
\end{equation}
where $g^{(1)}(\tau)$ is the first-order correlation function and corresponds to the Fourier tranform of the optical spectrum.
On the other hand, above threshold, the presence of photon population cycles in nanolasers causes the value of $g^{(2)}(0)$ to be larger than 1, in contrast to traditional lasers which display $g^{(2)}(\tau)=1$ for all time intervals $\tau$. 
At finite time-intervals, $g^{(2)}(\tau)$ tends towards the value of 1, in a damped oscillation fashion dropping periodically below 1 for a few cycles (see Fig.\ \ref{fig:g2} black line).
This feature arises from the presence of population cycles in nanolasers (even under constant pumping) and is a deviation from the Siegert relation (Eq.\ (\theequation)), which does not pemit $g^{(2)}(\tau)$ values smaller than 1.

\section{Observation of population oscillations}

In a recent publication \cite{Takemura}, observation of a fast periodic modulation in the photon correlation $g^{(2)}(\tau)$ was reported for a single-mode vertical-cavity surface-emitting semiconductor laser displaying a spontaneous emission coupling factor of $\beta \approx 10^{-4}$ (as can be seen on Fig. 2 of Ref.\ \cite{Takemura}). The modulation was clearly observable at a pumping current of 2.47 mA, corresponding to a power of 1 \% above threshold and $\sim 150$ photons in the lasing mode. However, the modulation was weaker at a pumping current of 2.72 mA (11 \% above threshold and $\sim 1500$ photons in the lasing mode) and was below instrumental sensitivity at currents of 3.00 mA (23 \% above threshold and $\sim 2300$ photons in the lasing mode). 
It was attributed by the authors to relaxation oscillations excited by (generic) noise in the laser. 

\begin{figure}[!ht]
   \begin{center}
     \begin{tabular}{l}
   \includegraphics[width=7cm]{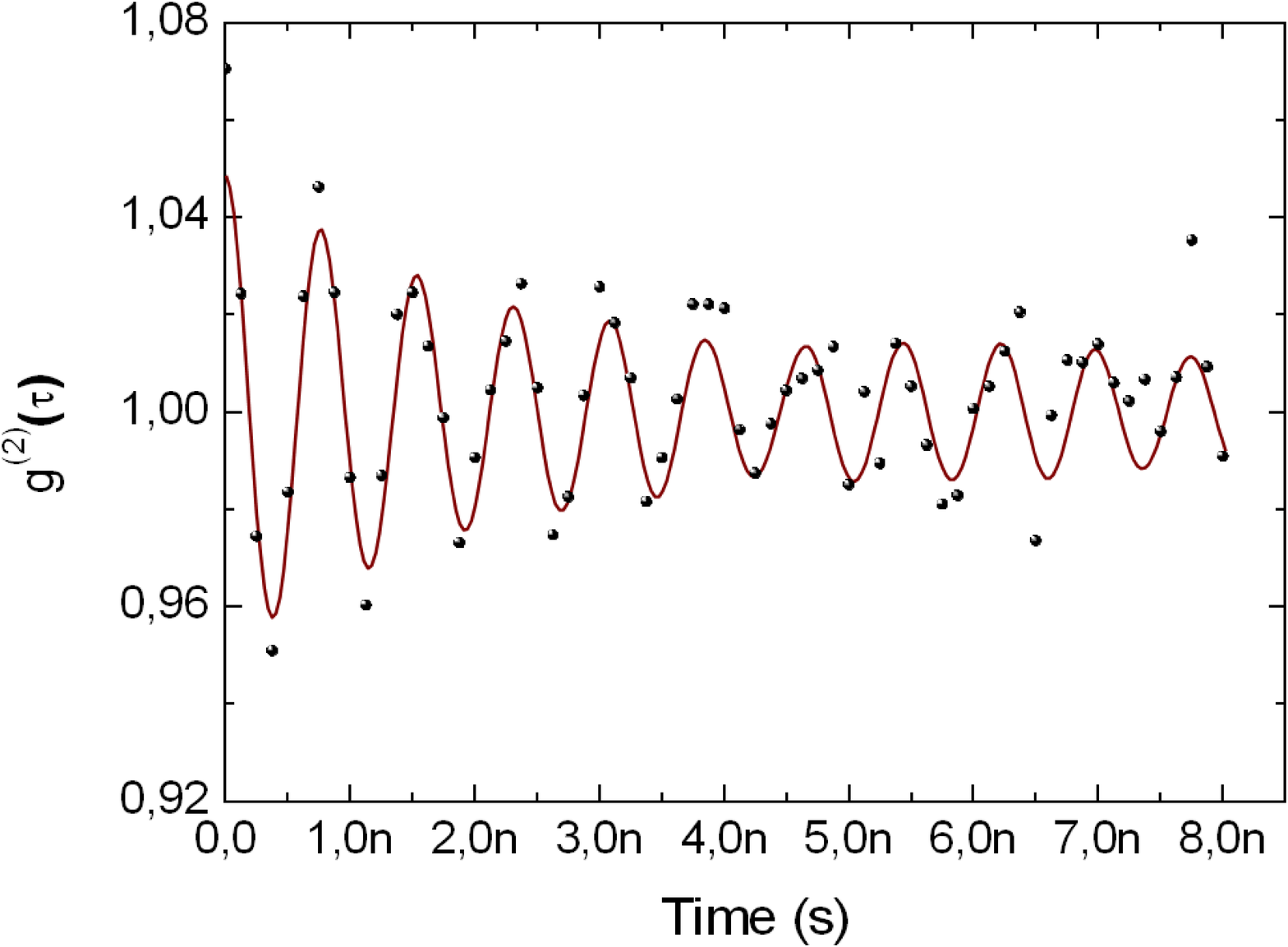} 
     \end{tabular}
   \end{center}
   \caption{ \label{fig:takemura} Second order autocorrelation function  $g^{(2)}(\tau)$ as function of time. Black dots:  experimental data from \cite{Takemura}. 
   Red continuous line: simulation through the discretization noise model.}
\end{figure}

We argue that in that experiment the oscillations are driven by discretization noise, as our model can fit the experimental second order autocorrelation function $g^{(2)}(\tau)$, with no adjustable parameters other than the cavity lifetime $1/\Gamma_c = 3.4$ ps and the spontaneous emission lifetime $1/\gamma_{\parallel} = 4$ ns extracted from \cite{Takemura}.
Figure \ref{fig:takemura} shows the experimental second order autocorrelation function $g^{(2)}(\tau)$ (black points) for a pump current of I = 2.6 mA (corresponding to a pumping power 6.5 \% above threshold and $\sim 1000$ photons in the cavity) and the fit using the discretization noise model is represented as a continuous red line. 

\section{Population oscillations and noise}

\begin{figure*}[!ht]
   \begin{center}
     \begin{center}
     
   \includegraphics[width=13.75cm]{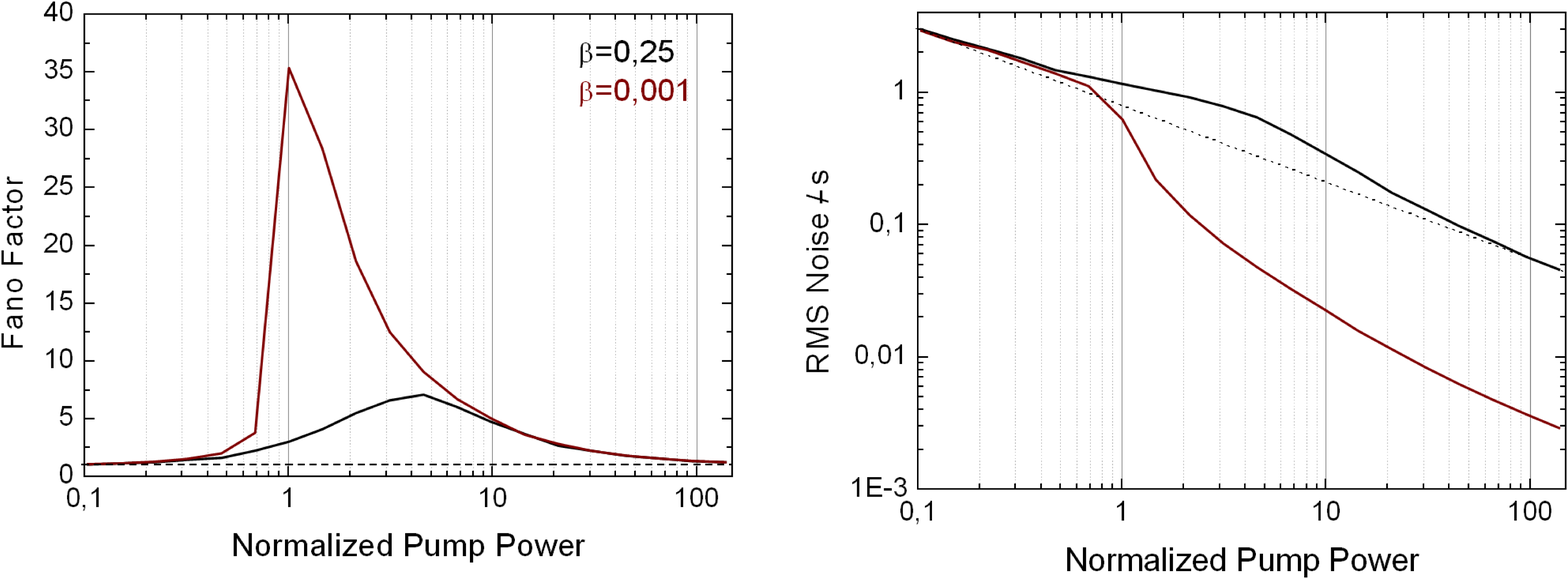} 
   
   \end{center}
   \end{center}
   \caption{ \label{fano} (Left) Power dependence of the Fano factor for $\beta =0.25$ (black line) and $\beta =0.001$ (red line).
   (Right) Power dependence of the normalized root-mean-square noise for $\beta =0.25$ (black line) and $\beta =0.001$ (red line).
   The dotted line (of logarithmic slope -1) is a guide to the eye.
   }
\end{figure*}

For small-$\beta$ lasers (in which the number of photons at threshold is large), or when the nanolaser response is averaged over times much longer than the relaxation oscillation period (and thus corresponds to an ensemble-averaged response), the population cycles arising from granularity appear as fluctuations around the (mean) steady-state populations.
These fluctuations are often characterized by the Fano factor \cite{Druten2000},
\begin{equation}
F =\frac{\sigma_s^2}{\bar{s}}=\bar{s} (g^{(2)}(0)-1) +1
\end{equation}
\noindent where $\sigma_s^2$ is the variance in the photon number of mean value $\bar{s}$. 
By analogy to phase transitions, it is often expected that fluctuations should be maximum at threshold \cite{Rice,Jin}, as can be seen in Fig.\ \ref{fano} for a low-$\beta$ laser. 
However, for class-B lasers ($\Gamma_c > \gamma_\parallel$) with large $\beta$, the Fano factor maximum occurs at powers significantly higher than the stimulated emission threshold (see Fig.\ \ref{fano} left), as shown by van Druten {\it et al.} \cite{Druten2000}. 
While this discrepancy between the laser threshold and the fluctuation maximum has been interpreted as an indication that such lasers produce partially chaotic light \cite{Hofmann2000} and become coherent only at pumping powers much higher than that of the Fano maximum, our results indicate that the increase of the Fano factor not due to the presence of a chaotic component in the laser output but arises from the large amplitude of the photon population cycles that are excited and sustained by discretization noise.

An alternative designation of the magnitude of the fluctuations, useful in engineering, is the normalized root-mean-square noise (nRMS), which is related to $g^{(2)}(0)$ by
\begin{equation}
nRMS =\frac{\sqrt{\sigma_s^2}}{\bar{s}}=\sqrt{ g^{(2)}(0)-1 +\frac{1}{\bar{s}}}
\end{equation}
\noindent As can be seen on Fig.\ \ref{fano} (right), for traditional low-$\beta$ lasers and above threshold, the nRMS noise drops rapidly to very small values. As the signal-to-noise ratio is essentially the inverse of nRMS, this feature makes the output of low-$\beta$ lasers an ideal carrier for high-fidelity information transmission.
For high-$\beta$ nanolasers, on the other hand, the nRMS noise remains quite high even at pumping powers much above threshold, both because of the presence of population cycles (which cause a large value for $g^{(2)}(0)$) and the relatively small value of $\bar{s}$. 
As the signal-to-noise ratio is of the order of 1 even high above threshold, this implies that such lasers are ill-adapted to the transmission of information, unless particular noise-reducing protocols are implemented taking into account the population cycles. 

\section{Conclusion}

In conclusion, the output of a continuously-pumped high-$\beta$ nanolaser cannot obey the traditional (continuous, real-number) laser equations, because of the small numbers of intra-cavity photons and dipoles involved, which bring forth the discrete nature of these small populations.
Such nanolasers should display population cycles both for the intra-cavity photons and the excited dipoles because of the intrinsic stochasticity of the photon and dipole granularity, analogous to what is observed in small ecosystems. 
These cycles can be understood also as resulting from relaxation oscillations driven synchronously by the sudden and discrete jumps in photon and dipole numbers, due to the elementary emission processes. 
The photon population cycles are observable as a periodic modulation of the second-order correlation function of the laser, as was recently reported by Takemura {\it et al.} \cite{Takemura}.  
The presence of population cycles in high-$\beta$ nanolasers may appear as excess noise which may hinder information processing and transmission with such lasers.
\\

The authors acknowledge numerous fruitful discussions with Dr.\ S. Barbay and Dr.\ G.-L. Lippi, and financial support from the C'NANO APOLLON project and from the French National Research Agency (ANR) through the Nanoscience and Nanotechnology Program (project NATIF ANR-09-NANO-P103-36).


\begin{thebibliography}{99}

\bibitem{Strauf2006} 
S.\ Strauf, K.\ Hennessy, M.\ T.\ Rakher, Y.-S.\ Choi, A.\ Badolato, L.\ C.\ Andreani, E.\ L.\ Hu, P.\ M.\ Petroff, and D.\ Bouwmeester, 
Phys.\ Rev.\ Lett.\ \textbf{96}, 127404 (2006).

\bibitem{Ulrich2007} 
S.\ M.\ Ulrich, C.\ Gies, S.\ Ates, J.\ Wiersig, S.\ Reitzenstein, C.\ Hofmann, A.\ L\"offler, A.\ Forchel, F.\ Jahnke, and P.\ Michler, 
Phys.\ Rev.\ Lett.\ \textbf{98}, 043906 (2007).

\bibitem{Hostein2010} 
R.\ Hostein, R.\ Braive, L.\ Le Gratiet, A.\ Talneau, G.\ Beaudoin, I.\ Robert-Philip, I.\ Sagnes, and A.\ Beveratos, 
Opt.\ Lett.\ \textbf{35}, 1154 (2010).

\bibitem{Noginov2009} 
M.A.\ Noginov, \textit{et al.}, 
Nature \textbf{460}, 1110 (2009).

\bibitem{Hill2009} 
M.T.\ Hill, \textit{et al.}, 
Opt.\ Exp.\ \textbf{17}, 11107 (2009).

\bibitem{Nezhad2010} 
M.P.\ Nezhad, \textit{et al.}, 
Nature Photonics \textbf{4}, 395 (2010).

\bibitem{Rice} 
P.R.\ Rice and H.J.\ Carmichael, 
Phys.\ Rev.\ A {\bf 50}, 4318 (1994).

\bibitem{Choudhury} 
K.\ Roy-Choudhury, S.\ Haas and A.F.J.\ Levi, Phys.\ Rev.\ Lett.\ \textbf{102}, 053902 (2009); K.\ Roy-Choudhury and A.F.J.\ Levi, 
Phys.\ Rev.\ A \textbf{81}, 013827 (2010).

\bibitem{Elvira} 
D.\ Elvira, X.\ Hachair, V.B.\ Verma, R.\ Braive, G.\ Beaudoin, I.\ Robert-Philip, I.\ Sagnes, B.\ Baek, S.W.\ Nam, E.A.\ Dauler, I.\ Abram, M.J.\ Stevens and A.\ Beveratos, 
Phys.\ Rev.\ A \textbf{84}, 061802(R) (2011).

\bibitem{Gardiner}
C.\ Gardiner and P.\ Zoller, 
{\it Quantum noise} (Springer, 2004).

\bibitem{Haroche}
J.M.\ Raimond, P.\ Goy, M.\ Gross, C.\ Fabre, and S.\ Haroche, 
Phys.\ Rev.\ Lett.\ \textbf{49}, 1924 (1982).

\bibitem{Walther}
D.\ Meschede, H.\ Walther, and G.\ Mueller, 
Phys.\ Rev.\ Lett.\ \textbf{54}, 551 (1985).

\bibitem{Kimble}
J.\ McKeever, A.\ Boca, A.D.\ Boozer, J.R.\ Buck, and H.J.\ Kimble, 
Nature \textbf{425}, 268 (2003).


\bibitem{Winger}
M.\ Winger, T.\ Volz, G.\ Tarel, S.\ Portolan, A.\ Badolato, K.\ J.\ Hennessy, E.L.\ Hu, A.\ Beveratos, J.\ Finley, V.\ Savona and A.\ Imamoglu,
Phys.\ Rev.\ Lett.\ \textbf{103}, 207403 (2009)

\bibitem{McKane2005} 
A.J.\ McKane and T.J.\ Newman, 
Phys.\ Rev.\ Lett.\ \textbf{94}, 218102 (2005).

\bibitem{Aparicio2001} J.P.\ Aparicio and H.G.\ Solari, Math.\ Bio.\ \textbf{169}, 15 (2001).

\bibitem{Takemura} 
N.\ Takemura, J.\ Omachi and M.\ Kuwata-Gonokami, 
Phys.\ Rev.\ A \textbf{85}, 053811 (2012).

\bibitem{Druten2000} 
N.J.\ van Druten, Y.\ Lien, C.\ Serrat, S.S.R.\ Oemrawsingh, M.P.\ van Exter, and J.P.\ Woerdman, 
Phys.\ Rev.\ A \textbf{62}, 053808 (2000).

\bibitem{Henry} 
C.H.\ Henry, IEEE J.\ Quantum Electronics \textbf{18}, 259 (1982); 
C.H.\ Henry, IEEE J.\ Quantum Electronics \textbf{19}, 1391 (1983).

\bibitem{Jin} 
R.\ Jin, \textit{et al.}, 
Phys.\ Rev.\ A \textbf{49}, 4038 (1994).

\bibitem{Hofmann2000} 
H.F.\ Hofmann and O.\ Hess,
Phys.\ Rev.\ A \textbf{62}, 063807 (2000).








\end{thebibliography}
\end{document}